\documentclass[pra,twocolumn,nofootinbib,floatfix,10pt]{revtex4-2}

\usepackage{amsmath}
\usepackage{amssymb}
\usepackage{wasysym}
\usepackage{graphicx}
\usepackage{color,soul}
\usepackage{physics}
\usepackage{siunitx}
\usepackage{dsfont}
\usepackage{float}
\usepackage[english]{babel}
\usepackage{blindtext}
\usepackage[english,nomargin,inline,marginclue,draft]{fixme}
\pdfpageheight\paperheight
\pdfpagewidth\paperwidth

\usepackage[colorlinks,linkcolor=blue,anchorcolor=blue,citecolor=blue,urlcolor=blue]{hyperref}

\fxusetheme{colorsig}
\FXRegisterAuthor{cg}{acg}{CG}  
\FXRegisterAuthor{th}{ath}{\color{blue}TH}  
\FXRegisterAuthor{ib}{aib}{\color{red}IB} 
\FXRegisterAuthor{sh}{ash}{\color{cyan}SH} 
\FXRegisterAuthor{db}{adb}{\color{green}DB} 
\FXRegisterAuthor{ps}{aps}{PS}
\makeatletter
\renewcommand*\FXLayoutInline[3]{%
  {\@fxuseface{inline}\ignorespaces{\color{fx#1}[#3: #2]}}}
\makeatother

\long\def\symbolfootnote[#1]#2{\begingroup%
\def\thefootnote{\fnsymbol{footnote}}\footnotetext[#1]{#2}\endgroup}

\def\nobreakbefore{%
  \relax\ifvmode\else
    \ifhmode
      \ifdim\lastskip > 0pt\relax
        \unskip\nobreakspace
      \else 
        \nobreakspace
      \fi
    \fi
  \fi
}
\let\oldcite\cite
\renewcommand\cite{\nobreakbefore\oldcite}

\begin{document}
\title{Higher-order and fractional discrete time crystals in Floquet-driven Rydberg atoms}

\author{Bang Liu$^{1,2}$}
\author{Li-Hua Zhang$^{1,2}$}
\author{Qi-Feng Wang$^{1,2}$}
\author{Yu Ma$^{1,2}$}
\author{Tian-Yu Han$^{1,2}$}
\author{Jun Zhang$^{1,2}$}
\author{Zheng-Yuan Zhang$^{1,2}$}
\author{Shi-Yao Shao$^{1,2}$}
\author{Qing Li$^{1,2}$}
\author{Han-Chao Chen$^{1,2}$}
\author{Bao-Sen Shi$^{1,2}$}
\author{Dong-Sheng Ding$^{1,2,\textcolor{blue}{\dag}}$}

\affiliation{$^1$Key Laboratory of Quantum Information, University of Science and Technology of China; Hefei, Anhui 230026, China.}
\affiliation{$^2$Synergetic Innovation Center of Quantum Information and Quantum Physics, University of Science and Technology of China; Hefei, Anhui 230026, China.}

\symbolfootnote[2]{dds@ustc.edu.cn}

\maketitle

\textbf{Higher-order and fractional discrete time crystals (DTCs) are exotic phases of matter where the discrete time translation symmetry is broken into higher-order and non-integer category. Generation of these unique DTCs has been widely studied theoretically in different systems. However, no current experimental methods can probe these higher-order and fractional DTCs in any quantum many-body systems. We demonstrate an experimental approach to observe higher-order and fractional DTCs in Floquet-driven Rydberg atomic gases. We have discovered multiple $n$-DTCs with integer values of $n$ = 2, 3, and 4, and others ranging up to 14, along with fractional $n$-DTCs with $n$ values beyond the integers. The system response can transition between adjacent integer DTCs, during which the fractional DTCs are investigated. Study of higher-order and fractional DTCs expands fundamental knowledge of non-equilibrium dynamics and is promising for discovery of more complex temporal symmetries beyond the single discrete time translation symmetry.}

\section*{INTRODUCTION}

Discovery and exploration of the phenomenon of spontaneous breaking of the discrete time translation symmetry, which results in realization of a counterintuitive phase of matter that was initially proposed by Wilczek \cite{wilczek2012quantum}, have sparked enormous interest in the condensed matter physics field. These elusive structures exhibit recurring patterns over time, thus providing a platform to study the distinctive temporal symmetry breaking in different systems both experimentally and theoretically \cite{zhang2017observation,choi2017observation,watanabe2015absence,syrwid2017time,huang2018clean,gong2018discrete,yao2020classical,li2012space,else2016floquet,PhysRevA.91.033617,autti2018observation,smits2018observation,pizzi2021bistability,autti2021ac,trager2021real}. The response of the system in time is crystallized exhibiting period doubling effect \cite{zhang2017observation,choi2017observation}, and not for the time itself. In general, time crystals can be divided into discrete and continuous time crystals \cite{sacha2017time,else2020discrete,kongkhambut2022observation,zaletel2023colloquium,sacha2020time} that exhibit discrete and continuous time-translation symmetry breaking, respectively. In addition, stabilized dissipative time crystals \cite{kessler2021observation} and prethermal discrete time crystals (DTCs) \cite{vu2023dissipative,kyprianidis2021observation} in driven and dissipative systems have been reported. However, the emergent symmetry is not restricted to a single time translation symmetry \cite{zaletel2023colloquium}, and the different subharmonics are thus nontrivial and correspond to higher-order integer DTCs and fractional DTCs, which have been widely studied theoretically \cite{giergiel2018time, giergiel2019discrete, surace2019floquet, pizzi2019period, PhysRevLett.127.140603, pizzi2021classical, kelly2021stroboscopic, pizzi2021higher}. Similar to the integer DTC, a fractional DTC is an exotic phase of matter that exhibits time-translation symmetry breaking in the form of fractionalization. In a fractional time crystal, the system's behavior is repeated periodically, and not only in discrete time steps but also with a fractional multiple of the driving period proposed in the literature \cite{matus2019fractional,pizzi2021classical,pizzi2021higher}. 

The large dipole moment of the Rydberg atom allows researchers to build a quantum many-body system that can be used to study many-body dynamics and quantum scars \cite{bernien2017probing,keesling2019quantum,serbyn2021quantum,bluvstein_controlling_2021}, non-equilibrium phase transition, and self-organized criticality \cite{lee2012collective,carr2013nonequilibrium,ding2019Phase,helmrich2020signatures,klocke2021hydrodynamic,ding2022enhanced}. In addition, a DTC can be realized based on the emergence of a metastable regime in an open Rydberg atom system \cite{gambetta2019discrete}. In the open Rydberg atom system, the response of the Rydberg atom population is repeated periodically and breaks ergodicity without periodic driving, thus obeying a limit cycle regime \cite{ding2023ergodicity, wadenpfuhl2023emergence} and resulting in a continuous time crystal \cite{wu2023observation}. These differ from a DTC phase by the use of Floquet driving. Periodic Floquet driving and the long-range interactions between Rydberg atoms will break the system’s equilibrium; the time translation symmetry breaking will thus emerge as a complex process, providing a platform to observe the higher-order and fractional DTCs.

In this work, we experimentally observe higher-order and fractional DTCs in strongly interacting Rydberg atoms under external radio-frequency (RF) field periodic driving conditions. The periodic pulses effectively drive the system out of equilibrium, thus leading to complex subharmonic responses. The response periodicity is an integral multiple of the driving period with $n\geq2$ [even beyond period doubling]. The higher-order DTCs observed remain rigid with respect to subtle detuning changes. We also observed the phase transition between the second-order DTC (2-DTC) and the third-order DTC (3-DTC). In addition, between adjacent integer higher-order DTCs, the probe transmission can oscillate within a fractional driving period, which means that the system's response is repeated periodically but is not limited to integer driving periods. The experimental results show good agreement with theoretical predictions. The findings for the higher-order and fractional DTCs realized here could highlight the rich dynamics that can emerge from driven and dissipative systems and open up new avenues for exploration of non-equilibrium physics.

\section*{RESULTS}
\subsection*{Physical model and experimental diagram}
\begin{figure*}
\centering
\includegraphics[width=2.05\columnwidth]{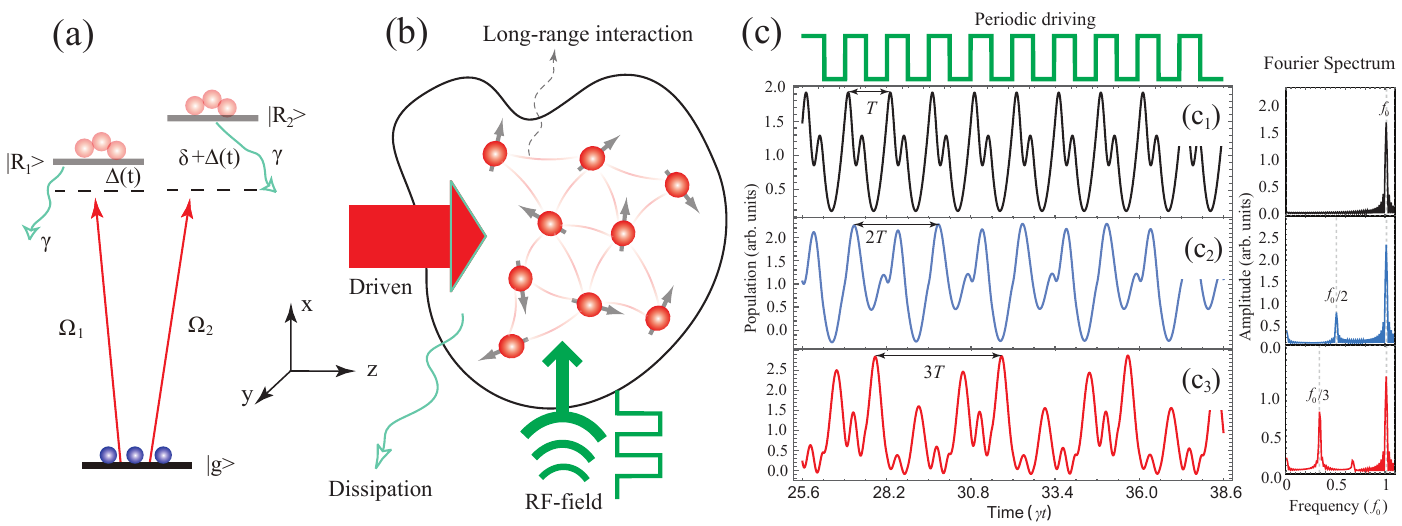}\\
\caption{\textbf{Physical model of DTC.} (a) Energy level diagram, which includes a ground state $\ket{g}$ and Rydberg states $\ket{R_1}$ and $\ket{R_2}$ (with a decay rate of $\gamma$). A laser drives the ground state $\ket{g}$ and Rydberg states $\ket{R_1}$ and $\ket{R_2}$ with respective detuning $\Delta(t)$ and $\delta+\Delta(t)$, in which $\delta$ is the interval between $\ket{R_1}$ and $\ket{R_2}$. $\Delta(t)$ is the detuning between laser beam and the Rydberg state $\ket{R_1}$, which is modified by an external pulsed radio-frequency (RF) field. (b) Physical diagram containing driven and dissipation Rydberg atoms and the driving RF field. The red transparent lines between Rydberg atoms indicate the long-range interactions. (c) Theoretical simulation of DTCs. The square wave-shaped line shows the time sequence for the external field. The left column (c1)-(c3) represents the time flows of the Rydberg atom population $\rho_{R_2R_2}$ under various conditions: [$V$ =0 (with no interaction), $\Omega_{1,2}$ = 4.49$\gamma$, $\Delta$ = 0, $\Delta_0$ = -12.18$\gamma$, $\delta$ = 10.34$\gamma$ for the black line], [$V$ = -20.51$\gamma$, $\Omega_{1,2}$ = 4.49$\gamma$, $\Delta$ = 0, $\Delta_0$ = -12.18$\gamma$, $\delta$ = 10.34$\gamma$ for the blue line], and [$V$ = -20.51$\gamma$, $\Omega_{1,2}$ = 4.49$\gamma$, $\Delta$ = 0, $\Delta_0$ = -17.95$\gamma$, $\delta$ = 10.34$\gamma$ for the red line]. The right column images show the corresponding Fourier spectra. During this process, the driving frequency $f_0$ = 1.28$\gamma$.}
\label{setup}
\end{figure*}

To observe breaking of the time translation symmetry, we constructed an experimental platform that combined a quantum many-body system with periodic Floquet driving. This many-body system includes $N$ interacting two-level cesium atoms with a ground state $\ket{g}$ and Rydberg states with sublevels $\ket{R_1}$ and $\ket{R_2}$ (with decay rate $\gamma$). In the model, we apply square wave modulation to the detuning $\Delta(t) = \Delta_0 + \Delta$ when $0\leq t\text{ }\textless \text{ }T/2$, and $\Delta(t) = \Delta$ when $T/2 \leq t \text{ }\textless \text{ }T$. This can be realized by applying the pulsed RF field in further experiments, where $\Delta$ represents the detuning of the laser beam and $\Delta_0$ is the RF field-induced energy shift; further details are provided in Methods section. The lasers drive the atoms along the $z$ coordinate direction and the pulsed RF field illuminates the atoms along the $y$ direction, as shown in Fig. \ref{setup}(b). 

The Hamiltonian of system is based on periodically driving double Rydberg state model \cite{wu2023observation}: 
\begin{equation}
\begin{aligned}
    \hat{H}(t) & =\frac{1}{2}\sum_{i}\left(\Omega_{1}\sigma_{i}^{gR_1}+\Omega_{2}\sigma_{i}^{gR_2}+h.c.\right)\\ &-\sum_{i}\left(\Delta(t) n_{i}^{R_1}+(\Delta(t) +\delta)n_{i}^{R_2}\right) \\ &+\sum_{i\neq j}V_{ij}\bigg[n_{i}^{R_1}n_{j}^{R_2}+\frac{1}{2}(n_{i}^{R_1}n_{j}^{R_1}+n_{i}^{R_2}n_{j}^{R_2})\bigg]
\end{aligned}
\end{equation}
where $\sigma_{i}^{gr}$ ($r={R_1,R_2}$) represents the $i$-th atom transition between the ground state $\left| g \right\rangle$ and the Rydberg state $\left|  r \right\rangle$, $n_{i}^{R_1,R_2}$ are the population operators for the two Rydberg energy levels $\left|  R_1 \right\rangle$, and $\left|  R_2 \right\rangle$, and $V_{ij}$ are the interactions between the Rydberg atoms located in $\mathbf{r}_i$ and $\mathbf{r}_j$ [through the van der Waals interaction $V_{ij} = C_6 /\left|\mathbf{r}_i-\mathbf{r}_j\right|^6$]. The Lindblad jump terms are given by $\mathcal{L}_r = (\gamma_{r}/2) \sum_i (\hat{\sigma}_i^{r g} \hat{\rho} \hat{\sigma}^{ gr}_i - \{\hat{n}_i^{r},\hat{\rho}\})$, which represents the decay process from the Rydberg state $\left| r \right\rangle$  ($r={R_1,R_2}$) to the ground state $\left| g \right\rangle$. Because of the periodic feature of $\Delta(t)$, the Hamiltonian of the system is symmetrical in the discrete time translation, in which $\hat{H}(t)$ = $\hat{H}(t+nT)$ ($n\in\mathbb{Z}$). Using the mean-field treatment, we calculate the master equation $\partial_t \hat{\rho} = i [\hat{H},\hat{\rho}] + \mathcal{L}_{R_1}[\hat{\rho}] + \mathcal{L}_{R_2}[\hat{\rho}]$ and obtain the matrix elements for $\rho_{R_1R_1}(t)$ and $\rho_{R_2R_2}(t)$, see more details in Method Sections. This master equation, which includes the driven and dissipation system, many-body interactions, and Floquet driving, provides a way to determine the appearance of the non-equilibrium dynamics and the emergence of time translation symmetry breaking.The subharmonic response of the Rydberg atoms occurs because of the presence of the interaction between Rydberg atoms and the external RF field periodic driving; this breaks the symmetry of time translation. Under specific conditions, the system will enter into distinct stationary states that evolve periodically over a long period of time, and it also exhibits $\mathbb{Z}_n$ symmetry ($n\geq2$) that is protected by Floquet driving \cite{zaletel2023colloquium}.

Here, we show the simulated results for $\rho_{R_2R_2}(t)$ under the periodic modulation $\Delta (t)$ to simplify the explanation of the subharmonic responses, as shown in Fig. \ref{setup}(c). Figure \ref{setup}(c1) represents the population without interactions, where the population $\rho_{R_2R_2}(t)$ behaves with a harmonic response to the driving signal. Figure \ref{setup}(c2) and (c3) show the results obtained when the interaction strength $V\neq 0$, where the population $\rho_{R_2R_2}(t)$ displays subharmonic responses, showing the response frequencies of $f_0/2$ and $f_0/3$ [where $f_0$ is the driving frequency], respectively, as illustrated by the Fourier spectra in the right column of Fig. \ref{setup}(c). In this process, the periodic driving signal and the interactions between the Rydberg atoms produce the DTCs as a result of time translation symmetry breaking. Additional simulated results for the distinct higher-order and fractional DTCs can be found in the Methods section.

In the experiments, we used an external pulsed RF field to drive the cesium Rydberg atoms. The external RF field with electric field component $E_{\rm{RF}}$ and a frequency of $\omega$ was modulated to produce a square wave. The RF field perturbs the Rydberg states and induces additional RF sidebands \cite{miller2016radio}, thus shifting the Rydberg energy levels. We applied a three-photon electromagnetically-induced transparency (EIT) scheme to prepare the Rydberg atoms [from the ground state $\ket{6S_{1/2}}$ to the Rydberg state $\ket{49P_{3/2}}$ using three lasers with wavelengths of 852 nm (probe), 1470 nm, and 780 nm], and measured the Rydberg atom population based on the transmission of the probe field \cite{zhangRydberg,liuHighly}, see more details in Methods section.

\subsection*{Higher-order DTCs}

\begin{figure*}
\centering
\includegraphics[width=2.1\columnwidth]{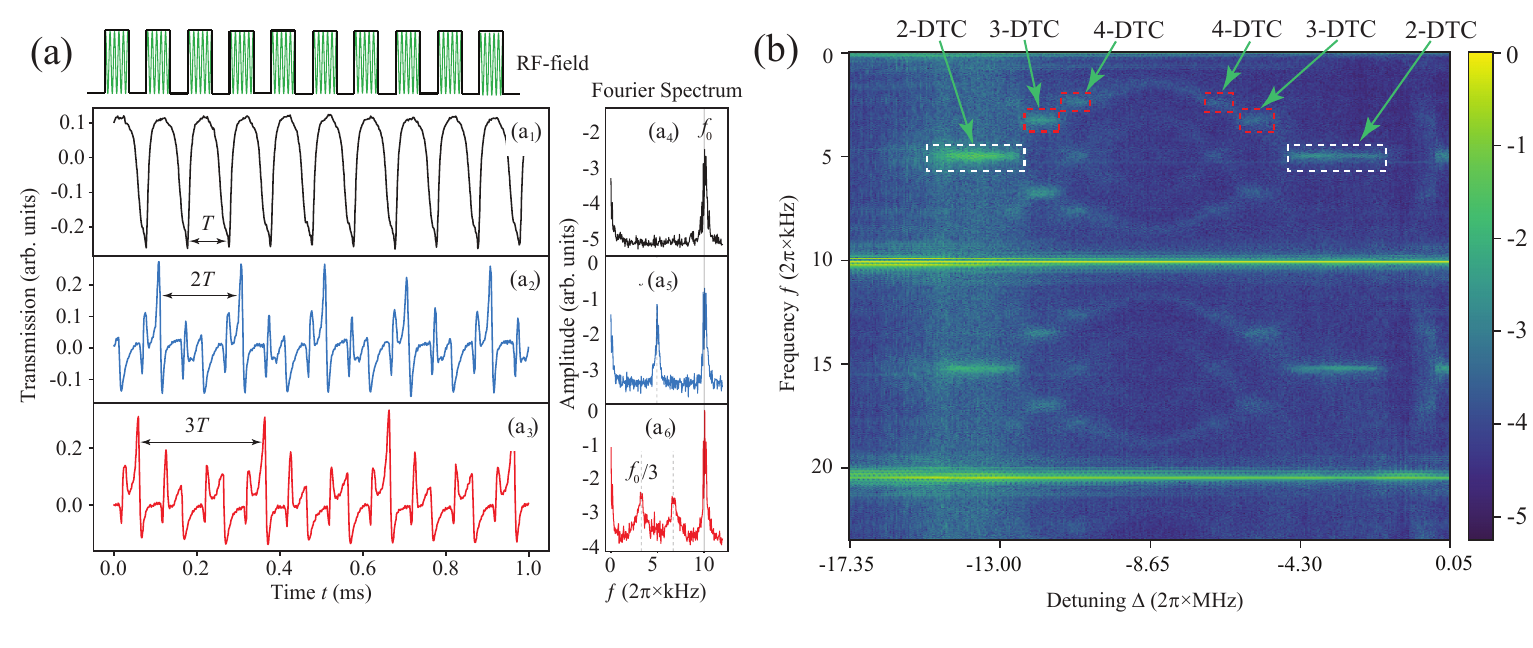}\\
\caption{\textbf{Observation of higher-order DTCs.} In the experiment, when we drive the system using a pulsed period $T$ of the RF field, we observe subharmonic responses that indicate that the probe transmission oscillates with a period that is an integer multiple of $T$. (a) Measured transmission of the system under different detunings of $\Delta$ = -$2\pi\times$24.25 MHz (a1), $\Delta$ = -$2\pi\times$14.37 MHz (a2), and $\Delta$ = -$2\pi\times$13.15 MHz (a3) [where the frequency offset here is induced by the distinctive $E_{\rm{RF}}$ in this dataset], which display different transmission pattern types. (a4) is the measured Fourier spectrum with no subharmonic response. (a5) and (a6) show subharmonic peaks that mark the $n$-DTCs with $n$ = 2 and 3, respectively. These spectra contain an integer multiple of $T$, thus revealing their distinct DTCs. The right column in (a) corresponds to the Fourier spectrum (with a logarithmic scale on the vertical axis), in which the blue and red datasets were measured during 10 multiple experimental trials. Here, the frequency of the RF field is $\omega = 2\pi\times 7.1$ MHz and its strength $E_{\rm{RF}}$ is 0.30 V/cm. (b) Measured phase map of the Fourier spectrum versus the detuning $\Delta$ under the driving field frequency condition of $f_0 = 2\pi\times 10$ kHz. The 2-DTC, 3-DTC, and 4-DTC are indicated by the dotted frames. The color bar represents the Fourier transform intensity. In these data sets, the frequency of the RF field is $\omega = 2\pi\times 7.1$ MHz and the electric field strength $E_{\rm{RF}}$ is 0.31 V/cm. The dataset presented here was measured from 10 multiple experimental trials. }
\label{experiment-high-order}
\end{figure*}

The observed subharmonic responses are unique features of DTCs that result from spontaneous breaking of the discrete time-translation symmetry. We recorded the time flows for probe transmission in the no-DTC, 2-DTC, and 3-DTC regimes when varying the laser detuning $\Delta$, as shown in Fig. \ref{experiment-high-order}(a). In Fig. \ref{experiment-high-order}(a1), the system's response shows harmonics with the same frequency as the driving RF field, and there are no peaks within the $0<f<2\pi\times$10 kHz section of the Fourier spectrum. In this case, the system is located far away from the resonance, the interactions between the Rydberg atoms are ignored, and thus the response is normal. However, when we change the system to the 2-DTC regime, in which the probe transmission oscillates with a periodicity of 2$T$ with respect to the driving field [see the blue line in Fig. \ref{experiment-high-order}(a2)], the period doubles. This period doubling effect is consistent with theoretical predictions presented in Ref. \cite{gambetta2019discrete}.

We also observed the phases of the higher-order $n$-DTCs with $n$ = 3 and 4, where these phases manifested as periodic oscillations with triple and quadruple values of $T$; the triple of $T$ can be found as indicated by the red line in Fig. \ref{experiment-high-order}(a3). The Fourier spectra of the oscillations in these scenarios demonstrate the appearance of the fractional peaks for $f_0$ with quantities of 1/2 and 1/3; see the right column in Fig. \ref{experiment-high-order}(b4-b6). The responses obtained with $n > 2$ reveal that the system evolves within the smaller discrete time-translation symmetry subgroup. The higher-order DTCs are a consequence of the interactions between the driven-dissipative many-body system and the periodic driving process.

These subharmonic oscillations arise due to many-body interactions, collective synchronization, and ergodicity breaking. The occurrence of these rigid and persistent subharmonic oscillations is a consequence of the intricate interactions between the Rydberg atoms in the system, leading to a breakdown of the dynamic equilibrium. The induced collective synchronization gives rise to a phenomenon called ``self-organization", where the Rydberg atoms are collectively excited and display coherent behavior on a macroscopic scale.

To map the full subharmonic responses of the system, we vary the laser detuning $\Delta$ and measure the Fourier spectrum under periodic RF field driving with a repetition frequency of $f_0 = 2\pi\times 10$ kHz (with a period of $T$), as depicted in Fig. \ref{experiment-high-order}(b). When we increased the laser detuning $\Delta$ from -$2\pi\times$17.35 MHz to -$2\pi\times$0.05 MHz, the system response showed complex non-equilibrium dynamics. In Fig. \ref{experiment-high-order}(b), the system's response shows oscillations with periods that are longer than that of the driving field. During this process, the system shows evolution with not only the doubled period $2T$ ($\mathbb{Z}_2$ symmetry), but also with $3T$ and $4T$, which exhibit $\mathbb{Z}_3$ and $\mathbb{Z}_4$ symmetries, respectively \cite{pizzi2021classical}. In addition, the higher-order $n$-DTCs with $n>2$ (indicated by the red dotted frame shown in Fig. \ref{experiment-high-order}(b)) are robust against the perturbations, as indicated by their stability when the detuning $\Delta$ was varied within a small range. When $\Delta$ is increased further, the $4$-DTC no longer becomes rigid to small changes in $\Delta$, and this results in a melting effect; see the Methods section for further explanation of these results.

\begin{figure*}
\centering
\includegraphics[width=2.08\columnwidth]{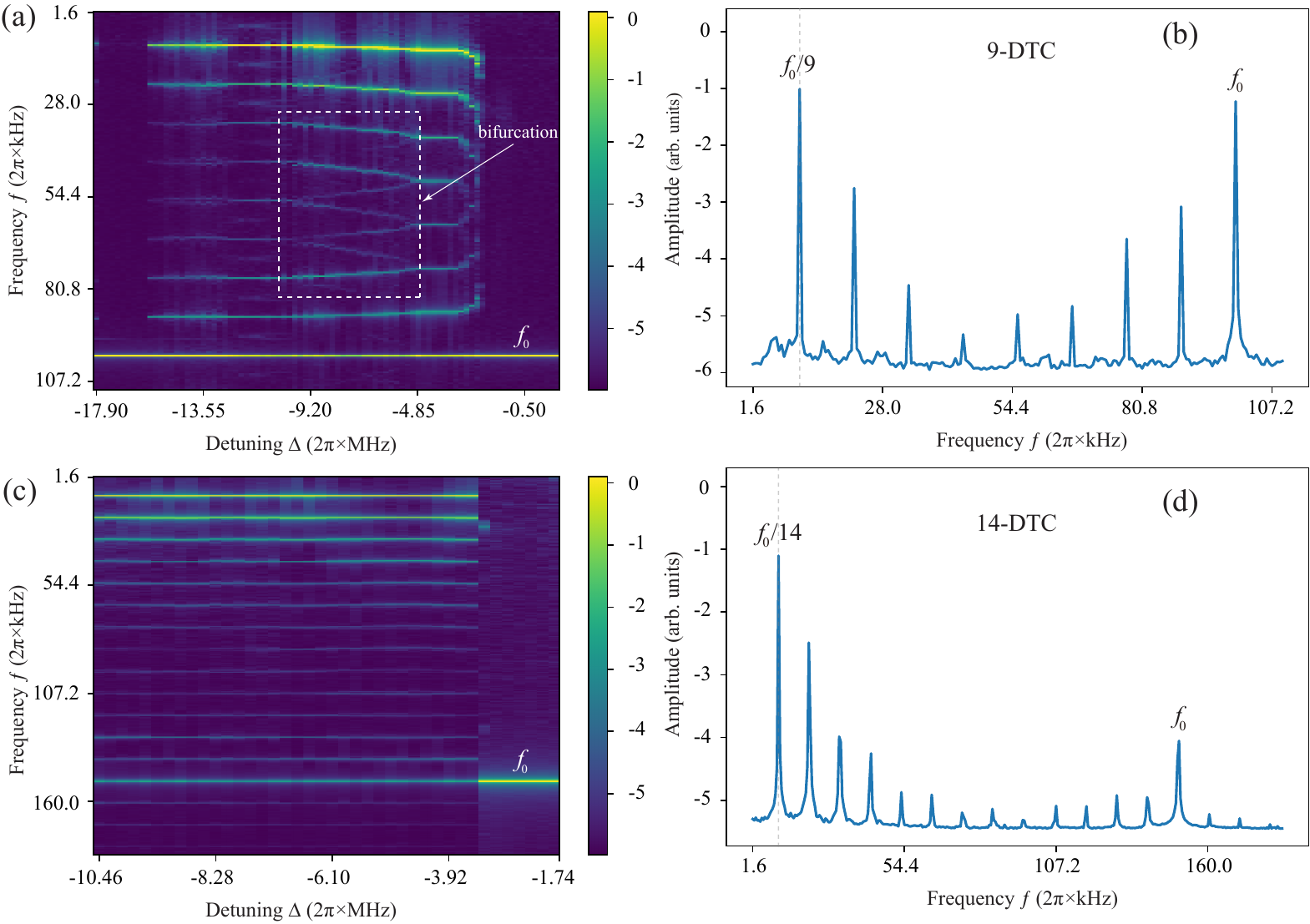}\\
\caption{\textbf{Higher-$n$ integer DTCs.} Measured phase maps of the Fourier spectrum at $f_0 = 2\pi\times 100$ kHz (a) and at $f_0 = 2\pi\times 150$ kHz (c). The frequency of the RF field was set at $\omega= 2\pi\times 7.1$ MHz in (a) and at $\omega= 2\pi\times 5.8$ MHz in (c). The electric field strength $E_{\rm{RF}}$ is 0.705 V/cm in (a) and 0.75 V/cm in (c). The dotted frame in (a) highlights a bifurcation effect. The color bar represents the Fourier transform intensity. (b) represents the Fourier spectrum when $\Delta= -2\pi\times 12.24$ MHz, which shows a 9-DTC signature. (d) presents the Fourier spectrum with $\Delta= -2\pi\times 6.11$ MHz, which corresponds to the features of a 14-DTC. The datasets presented here were measured from five multiple experimental trials for high visibility.}
\label{high-n}
\end{figure*}

In the system, increasing the driving frequency $f_0$ allowed us to observe the higher-$n$ integer DTCs. We increased the frequency $f_0$ to $f_0 = 2\pi\times$100 kHz and to $f_0 = 2\pi\times$150 kHz, and measured the phase maps of the Fourier spectrum, as shown in Fig. \ref{high-n}(a) and Fig. \ref{high-n}(c), respectively. Here, we selected a special detuning range $\delta\Delta$ in which higher-$n$ DTCs exist. In Figs. \ref{high-n}(a) and (c), we see comb-type structures in the Fourier spectrum when the $f_0$-pulsed driving field is applied. These comb lines correspond to the higher frequency orders of $n$-DTCs with $n$ = 8 and 9 in Fig. \ref{high-n}(a) and $n$ = 14 in Fig. \ref{high-n}(c), and the $\mathbb{Z}_8$, $\mathbb{Z}_9$, and $\mathbb{Z}_{14}$ symmetries arise. Figure \ref{high-n}(b) and (d) show the Fourier spectra at $\Delta= -2\pi\times 12.24$ MHz and $\Delta= -2\pi\times 6.11$ MHz, respectively. In these two figures, we can see that the first peaks occur at frequencies of $f_0/9$ and $f_0/14$, respectively, thus revealing their higher-order subharmonic responses to the driving field. Here, the corresponding first-peak frequencies of $f_0$/9 [Fig.\ref{high-n}(b)] and $f_0$/14 [Fig.\ref{high-n}(d)] are approximately same. This corresponds to the inherent oscillated property which is determined by dynamics near the limit cycle regime. These experimentally observed higher-order $n$-DTCs are consistent with the theoretical predictions presented in Ref. \cite{pizzi2021higher}.

\subsection*{Transition between distinct DTCs}
\begin{figure*}
\centering
\includegraphics[width=2.08\columnwidth]{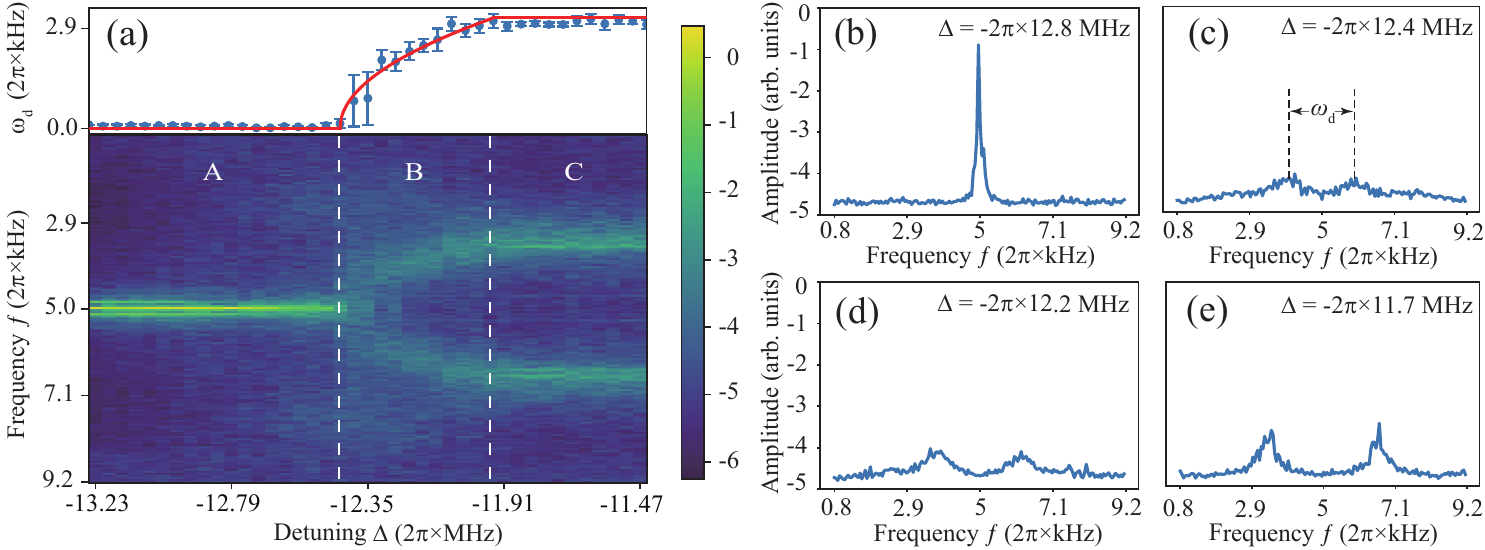}\\
\caption{\textbf{Phase transition.} (a) Phase maps of 2-DTC, 3-DTC, and their transitions. There are three regimes here, marked A, B, and C in lower plane of (a). In upper plane of (a), the red line is plotted using a fitting function of $\omega_d=\sqrt{(\Delta-\Delta_c)/a}$, where $a$ = 0.046 $\rm{Hz^{-1}}$ and $\Delta_c$ = -2$\pi\times$12.44 MHz. The blue data points are the average of five replicate measurements and the error bars represent the standard deviation. During this process, the driving frequency was set at $f_0 = 2\pi\times$10 kHz. The color bar represents the Fourier transform intensity. (b) Fourier spectrum at $\Delta$ = -2$\pi\times$12.8 MHz used to describe the 2-DTC phase. (c), (d) Fourier spectra of the intermediate states between 2-DTC and 3-DTC, where these spectra consist of a series of intermediate states. (e) shows the Fourier spectrum of the 3-DTC response. The distance between the two peaks in the transition regime is marked using $\omega_d$. The datasets presented here were measured from 50 multiple experimental trials to obtain high-resolution phase maps.}
\label{transition}
\end{figure*}

The observed higher-order $n$-DTCs allow us to demonstrate the occurrence of phase transitions between adjacent integer DTCs. These phase transitions are the result of breaking from the $\mathbb{Z}_2$ time translation symmetry to the $\mathbb{Z}_3$ symmetry. The upper panel of Fig. \ref{transition}(a) characterizes the measured frequency difference $\omega_d$ between splitted peaks.  The lower panel of Fig. \ref{transition}(a) shows the phase diagram of the Fourier spectrum acquired by scanning the laser detuning $\Delta$ with high resolution. In the lower panel of Fig. \ref{transition}(a), we see that there are three regimes [marked A, B, and C], including the 2-DTC regime, the 3-DTC regime, and their transition. We plotted the Fourier spectra for these regimes, as shown in Fig. \ref{transition}(b)-(e). Figure \ref{transition}(b) shows a Fourier spectrum at $\Delta = -2\pi\times$12.8 MHz, in which there is a single peak that corresponds to the 2-DTC. In Fig. \ref{transition}(e), the two peaks in the Fourier spectrum show the phase appearance of the 3-DTC.

The broken symmetry is characterized using a nonzero order parameter that is defined as the width between the two peaks $\omega_d = f_1-f_2$, where $f_1$ and $f_2$ are the frequencies corresponding to the two peaks. For the case where $\Delta<-2\pi\times 12.44$ MHz in regime A, the width $\omega_d$ is zero because the symmetry is not broken, and the system response remains stable in the 2-DTC phase. For the case where $\Delta\geq-2\pi\times 12.44$ MHz, the symmetry of the 2-DTC is broken, and the width $\omega_d$ then becomes nonzero as the two peaks appear at around $f = 2\pi\times$5 kHz with a $\Delta$-dependent width; see area B in the lower panel in Fig. \ref{transition}(a). During the phase transition process, the system response goes through a series of fractional responses with $2<n<3$. Therefore, we can map the entire process of breaking from the $\mathbb{Z}_2$ time translation symmetry to the $\mathbb{Z}_3$ symmetry. In Fig. \ref{transition}(c) and (d), we plotted two Fourier spectra at $\Delta = -2\pi\times 12.4$ MHz and $\Delta = -2\pi\times 12.2$ MHz, respectively, as examples. The measured widths $\omega_d$ = $2\pi\times 2.1$ kHz and $\omega_d$ = $2\pi\times 2.6$ kHz can be found in Fig. \ref{transition}(c)-(d). These two spectra show that the system is broken into a state with some type of non-integer time translation symmetry. These states are not robust to perturbations and therefore are not rigorous fractional DTCs \cite{pizzi2021higher}. When the system is tuned into area C, the system response then shows stability in the 3-DTC phase.

We also characterized the phase transition from regime A to regime B using the order parameter 
\begin{equation}
\omega_d=[(\Delta-\Delta_c)/a]^{\lambda}
\label{order parameter}
\end{equation}
Here, $\Delta_c = -2\pi\times 12.44$ MHz represents the critical point, $\lambda = 0.5$ is the fitted critical exponent, $a$ is a coefficient. While in regime A [$\Delta<\Delta_c = -2\pi\times 12.44$ MHz], $\omega_d=0$. The onset of the phase transition is characterized by symmetry breaking accompanied by a nonzero order parameter in regime B [as indicated by the red line in the upper panel of Fig. \ref{transition}(a)]. The continuity at $\Delta$ indicates that the system undergoes a second-order continuous phase transition. Furthermore, the system transits from the 2-DTC phase to the 4-DTC phase, where this transition is accompanied by the symmetry being broken two times within the detuning range from $-2\pi\times 14.0$ MHz$<\Delta<-2\pi\times 10.7$ MHz, as shown in Fig. \ref{experiment-high-order}(a).


\subsection*{Fractional DTCs}
\begin{figure*}
\centering
\includegraphics[width=2.08\columnwidth]{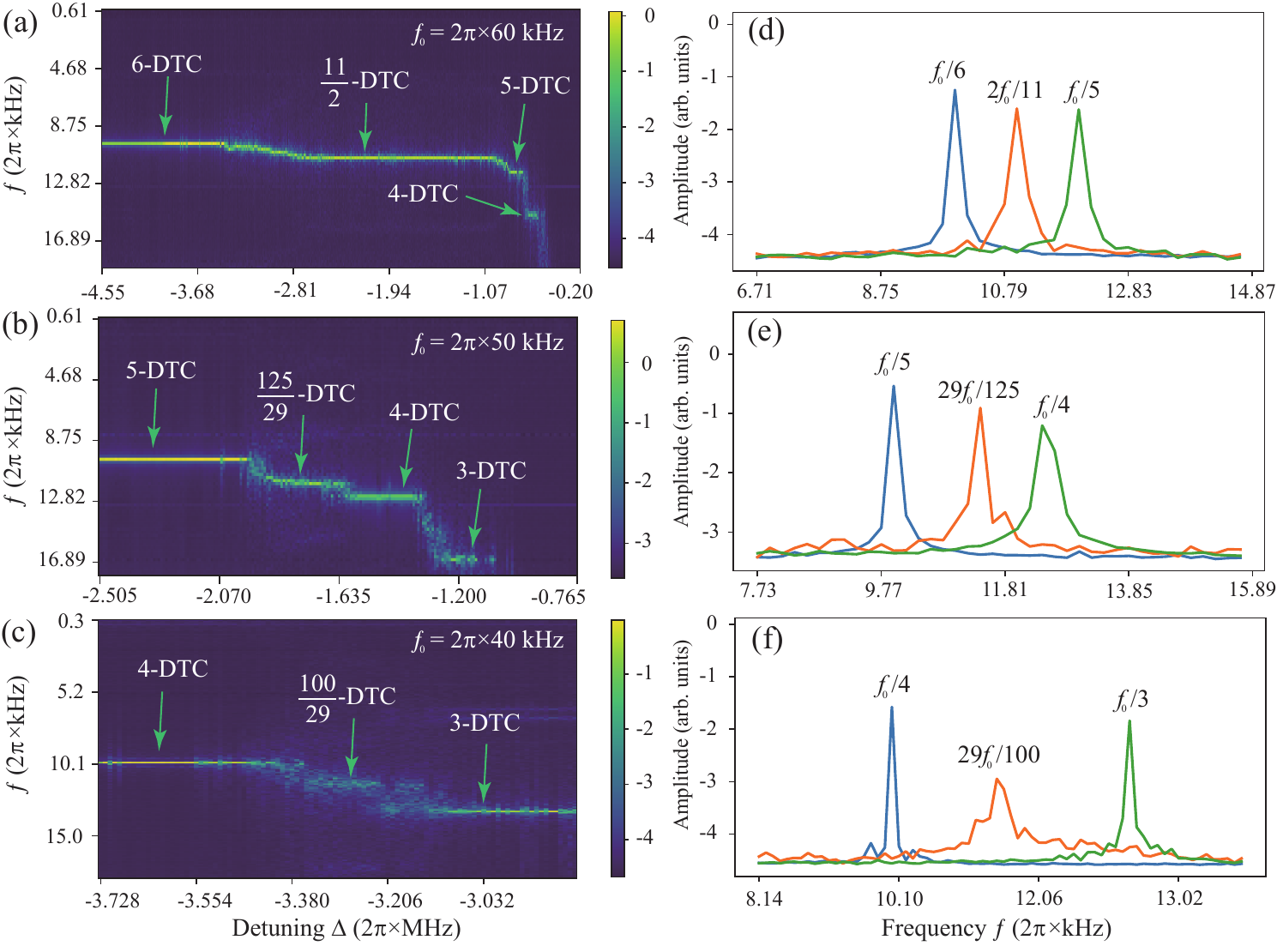}\\
\caption{\textbf{Fractional DTCs.} (a)-(c) Measured phase maps of Fourier spectra with different driving frequencies of $f_0$ = 2$\pi\times$40 kHz, $f_0$ = 2$\pi\times$50 kHz, and $f_0$ = 2$\pi\times$60 kHz, respectively. (d) Fourier spectra of the higher-order $n$-DTCs with integers $n$ = 6, 5, and of the fractional $n$-DTCs with $n$ = 11/2. (e) Fourier spectra of the higher-order $n$-DTCs with integers $n$ = 5, 4, and of the fractional $n$-DTCs with $n$ = 125/29. (f) Fourier spectra of the higher-order $n$-DTCs with integers $n$ = 3, 4, and of the fractional $n$-DTCs with $n$ = 100/29. The datasets presented here were measured from five multiple experimental trials for high visibility. Here, the fractional frequencies are the measured frequencies at the peak in Fourier spectrum. For instance, the value of 0.182\textit{f}\textsubscript{0} shown in (a) corresponds to 2/11\textit{f}\textsubscript{0}; the value of 0.232\textit{f}\textsubscript{0} in (b) corresponds to 29/125\textit{f}\textsubscript{0}; and the value of 0.29\textit{f}\textsubscript{0} in (c) corresponds to 29/100\textit{f}\textsubscript{0}. These values—0.182\textit{f}\textsubscript{0}, 0.232\textit{f}\textsubscript{0}, and 0.29\textit{f}\textsubscript{0}—are straightforward to be identified in the Fourier spectrum. }
\label{Fractional-DTC}
\end{figure*}

In this experiment, we also observed fractional $n$-DTCs, as predicted in Ref. \cite{pizzi2021higher}. These observations of the distinct higher-order $n$-DTCs allow us not only to investigate the phase transition, but also to study the characteristics of the intermediate states. We selected the values of $f_0$ = 2$\pi\times$40 kHz, $f_0$ = 2$\pi\times$50 kHz, and $f_0$ = 2$\pi\times$60 kHz to study the transition between the distinct integer higher-order $n$-DTCs. Figure \ref{Fractional-DTC}(a), (b), and (c) show the phase maps of the corresponding Fourier spectra, which were measured by scanning the laser detuning $\Delta$. From these results, we can see that frequency peaks occur between the adjacent integer $n$-DTCs, and these peaks remain stable over a certain detuning range $\delta\Delta$. For example, there is a peak between the 6-DTC and the 5-DTC, as shown in Fig. \ref{Fractional-DTC}(a), and similar results can be seen in Fig. \ref{Fractional-DTC}(b) and (c). These exotic peaks correspond to the fractional $n$-DTCs, with $n$ = $p/q$, where $p$ and $q$ are coprime integers. Figure \ref{Fractional-DTC}(d)-(f) show the corresponding Fourier spectra of the fractional DTCs. In addition to the adjacent integer $n$-DTCs, fractional $n$-DTCs with $n$ = 11/2, 125/29, and 100/29 were observed; see the orange lines in Fig. \ref{Fractional-DTC}(d)-(f).

Whether or not \textit{n} is integer and fractional is determined by the proportion in which the dynamical equilibrium is broken in the driving cycle. The observed fractional $n$-DTCs come from the process of symmetry breaking from high symmetry into fractional symmetry. In this process, the ergodicity of system is delayed \cite{zaletel2023colloquium} and a time average is considered equivalent to an ensemble average but only confined to a limited phase space. In this case, the system does not explore all possible states in phase space, and the Rydberg atoms become trapped in a cyclic excitation and decay in a fractional manner, exhibiting peculiar cluster excitations in spatial space. The observed fractional DTCs are robust with respect to the subtle variations in $\Delta$. This robustness is due to the emergence of the protected fractional symmetry created by a combination of many-body interactions and Floquet driving. This makes the fractional DTCs resistant to perturbations and ensures that they preserve their time-keeping behavior, and thus confirms the results of investigations of stable fractional DTCs presented in Ref. \cite{pizzi2021higher}.

In the experiments, it was not easy to observe the fractional DTCs between 2-DTC and 3-DTC because we can only see a smooth gradient in Fig. \ref{transition}(a). Because the rigid range of the fractional DTCs between 2-DTC and 3-DTC is relatively small, it is difficult to see a stable fractional DTC within the limited stability range of the system. However, the fractional DTCs between the $n$-DTCs with high integer values of $n$ are more stable than those with relatively lower values; as a result, we can observe a fractional DTC between $n$ = 5 and $n=6$ that is more stable than that observed between $n$ = 3 and $n=4$ [see Fig. \ref{Fractional-DTC}(a) and Fig. \ref{Fractional-DTC}(c)]. As our system is near the limit cycle regime, there is an inherent oscillated frequency range when no RF-field driving. Thus, the higher-order and fractional DTCs are easier to appear at large $f_0$.

\section*{DISCUSSION}
In the system, for lower values of $f_0$, the observed integer $n$-DTC ($n$ = 2 $\sim$ 4) remained stable against the subtle perturbations caused by detuning [which induces fluctuations in the Rydberg atom population]. As shown in Fig. \ref{experiment-high-order}(a), the detuning interval for the appearance of the 2-DTC is larger than that for both the 3-DTC and 4-DTC. In our system, the emergence of the discrete time translation symmetry occurs over a limited frequency range, and it is fragile at both lower and higher oscillation frequencies with respective to the driving frequency. In the theoretical model, all parameters can be controlled arbitrarily and there is no limitation for oscillation frequency of the DTC. However, in practical systems, several parameters are fixed; for instance, the interaction strength, the lifetime of Rydberg atoms, and the frequency interval between Rydberg sublevels. These parameters establish the boundary conditions of the Lindblad master equation, subsequently influencing the dynamical evolution conditions under which the transition from the normal regime to the time translation symmetry breaking regime occurs. We cannot locate visible stability within any detuning range for higher-order $n$-DTCs with $n>4$ at $f_0 = 2\pi\times 10$ kHz. However, higher integer $n$-DTCs with $n>4$ can be found more easily at higher values of $f_0$. The observed higher-order integer $n$-DTCs may have potential applications in quantum information science and technology because these DTCs provide a method to establish a versatile bridge for a frequency transducer from high frequency to low frequency. 

In summary, we have presented an experimental observation of higher-order and fractional DTCs in a periodic Floquet-driven Rydberg atomic gas. The interplay between the driving field and the interactions between the Rydberg atoms leads to the emergence of rich time translation symmetry breaking. Phase transitions between the adjacent integer DTCs were observed, where the discrete time translation symmetry of the system acts as a further breakage, displaying a full dynamics of symmetry breaking. These experimental results confirm previous theoretical predictions \cite{matus2019fractional, giergiel2018time, giergiel2019discrete, surace2019floquet, pizzi2019period, kelly2021stroboscopic, pizzi2021higher}. Investigation of the higher-order and fractional DTCs in the Rydberg atom system promotes the theory of the DTCs, and opens new avenues for exploration of the rich landscape of time translation symmetry breaking in driven quantum systems.

\section*{METHODS}
\subsection*{Experimental details}

The experiment is based on three-photon EIT scheme to prepare and measure the populations of Caesium Rydberg atoms. The laser system consists of the probe, dressing and coupling lasers. The probe laser (852 nm) is splitted into a reference and a probe, which are focused into a 7-cm glass cell with $1/e^2$-waist radius of approximately 200 $\mu$m and couples the ground state $\ket{6S_{1/2},F=4}$ to the intermediate state  $\ket{6P_{3/2},F = 5}$. The dressing laser (1470 nm) is focused into the cell with $1/e^2$-waist radius of approximately 500 $\mu$m and couples the two intermediate states $\ket{6P_{3/2},F = 5}$ and $\ket{7S_{1/2},F = 4}$. The coupling laser (780 nm, $1/e^2$-waist radius of approximately 500$\mu$m) drives the transition from$\ket{7S_{1/2},F = 4}$ to the Rydberg state $\ket{49P_{3/2},F = 4}$. The probe laser is parallel with the reference light, and counterpropagates with the dressing and the coupling lasers. The use of four-energy-level structure above 580 nm is to avoid the ionization shielding effect of Caesium atoms in the atomic vapor cell due to photoelectric ionization. The two 852 nm laser beams are received by a balanced photoelectric detector for differential amplification measurement. The glass cell is placed in the center of two parallel copper electrodes where the distance between the two parallel copper electrodes is 4 cm. An arbitrary function generator(Rigol DG4000 series) is connected to the electrode plates to generate the RF electric field required for the experiment.

\subsection*{Mean field approximation}
Because the number of Rydberg atoms $N$ in our experimental system is high and because of the thermal motion of the atoms, we can ignore the correlations between the atoms. Simultaneously, the interactions between the atoms can be treated using the mean-field approximation, and the many-body density matrix $\rho$ is decoupled into the tensor products of individual matrices. The mean field approximation is a valid option for dealing with the DTCs \cite{giergiel2019discrete}. In the mean field method, the mean values of the elements of the system density matrix $\rho$ are governed by the following equations of motion:

\begin{equation}
\begin{aligned}
\frac{\partial}{\partial t} \rho_{R_1 R_1} & =i \frac{\Omega}{2}\left(\rho_{g R_1}-\rho_{R_1 g}\right)-\gamma \rho_{R_1 R_1}, \\
\frac{\partial}{\partial t} \rho_{R_2 R_2} & =i \frac{\Omega}{2}\left(\rho_{g R_2}-\rho_{R_2 g}\right)-\gamma \rho_{R_2 R_2}, \\
\frac{\partial}{\partial t} \rho_{g R_1} & =i \frac{\Omega}{2}\left(\rho_{R_1 R_1}+\rho_{R_2 R_1}-\rho_{g g}\right) \\ 
& +i\left(\Delta(t)-V_{\rm{MF}}+i \frac{\gamma}{2}\right) \rho_{g R_1}, \\
\frac{\partial}{\partial t} \rho_{g R_2} & =i \frac{\Omega}{2}\left(\rho_{R_2 R_2}+\rho_{R_1 R_2}-\rho_{g g}\right)\\
& +i\left(\Delta(t)+\delta-V_{\rm{MF}}+i \frac{\gamma}{2}\right) \rho_{g R_2}, \\
\frac{\partial}{\partial t} \rho_{R_1 R_2} & =i \frac{\Omega}{2}\left(\rho_{g R_2}-\rho_{R_1 g}\right)-i\left(\delta-i \gamma\right) \rho_{R_1 R_2},
\end{aligned}
\end{equation}
where $V_{\rm{MF}} = V(\rho_{R_1 R_1}+\rho_{R_2 R_2})$ is the mean field shift, and we set the effective Rabi frequency $\Omega_1=\Omega_2=\Omega$ here. Here, we treat a simplified model by using same interaction $V_{ij}=V$ by ignoring the difference between different sublevels of Rydberg atoms. By solving the equations above, we can obtain the time response of the system and can also obtain the Fourier spectrum via discrete Fourier transformation. We look for the types of stationary state that evolve periodically over a long time period, and the evolution exhibits $\mathbb{Z}_n$ symmetry ($n\geq2$), where $n$ can be both an integer and a fraction. The results of the numerical calculations are in the Supplementary Note 1 and 2.

\section*{DATA AVAILABILITY}
The data generated in this study have been deposited in the Zenodo database (\href{https://doi.org/10.5281/zenodo.13183230}{https://doi.org/10.5281/zenodo.13183230}).

\section*{CODE AVAILABILITY}
The custom codes used to produce the results presented in this paper are available from the corresponding authors upon request.

\bibliography{ref}

\section*{Acknowledgements}
D.-S.D. thanks for the previous discussions with Prof. Igor Lesanovsky and Prof. C. Stuart Adams on time crystals. We acknowledge funding from the National Key R and D Program of China (Grant No. 2022YFA1404002), the National Natural Science Foundation of China (Grant Nos. U20A20218, 61525504, and 61435011), the Anhui Initiative in Quantum Information Technologies (Grant No. AHY020200), and the Major Science and Technology Projects in Anhui Province (Grant No. 202203a13010001).

\section*{Author contributions}
D.-S.D. conceived the idea for the study. B.L. conducted the physical experiments and developed the theoretical model. B.L. collected data with assistance of J.Z, Z.Y.Z., S.Y.S., Q.L. and H.C.C., discussed data with L.H.Z., Y.M., T.Y.H., Q.F.W. and B.S.S. The manuscript was written by D.-S.D and B.L. The research was supervised by D.-S.D. All authors contributed to discussions regarding the results and the analysis contained in the manuscript.
\\
\section*{Competing interests}
The authors declare no competing interests.

\section*{SUPPLEMENTARY NOTE 1: Integer DTCs: 2-DTC, higher-order DTCs}

By varying the parameters, the system can be tuned into the DTC regime in which the response of the Rydberg atom population is subharmonic to the driving frequency. By scanning the detuning $\Delta$ from -2.92$\gamma$ to -22.35$\gamma$, we plotted the phase map of the Fourier spectrum of the Rydberg atom populations $\rho_{R_1R_1}$ and $\rho_{R_2R_2}$; the results are shown in Supplementary Fig. \ref{phase for 2-DTC}(a) and Supplementary Fig. \ref{phase for 2-DTC}(b), respectively. There are subharmonic peaks in the Fourier spectrum, we find that both Rydberg atom populations, i.e., $\rho_{R_1R_1}$ and $\rho_{R_2R_2}$, go through the stable phases from 2-DTC to 4-DTC [exhibiting $\mathbb{Z}_2$ to $\mathbb{Z}_4$ symmetry]. The $n$-DTCs with $n>2$ (i.e., beyond period doubling) are called the higher-order DTCs. The results of these investigations of the higher-order DTCs are consistent with those of previous studies \cite{pizzi2021classical,pizzi2021higher}.

\section*{SUPPLEMENTARY NOTE 2: Bifurcation of 2-DTC and fractional DTCs}

Furthermore, we set the system into the bifurcation regime by changing the parameters $\delta$, $\Omega_{1,2}$, and $\Delta_0$, and then recording the response of the Rydberg atom population $\rho_{R_2R_2}$. The calculated results are presented in Supplementary Fig. \ref{phase for Fractional DTC}(a), which shows the phase diagram of the Fourier spectrum of $\rho_{R_2R_2}$. When $\Delta_0$ is increased from -7.04$\gamma$ to -9.36$\gamma$, the response is normally harmonic with respect to the driving frequency in the harmonic regime. By increasing $\Delta_0$ further, the response becomes subharmonic to the driving frequency, as indicated by the series of bright lines within the range $f<f_0$ in the phase diagram. In the subharmonic regime, we can observe numerous stationary oscillation states, which are manifested as frequency peaks of $2$-DTC, the higher-order 4-DTC, and a series of fractional DTCs.

When the system changes state between the distinct integer DTCs, this change is accompanied by discrete time translation symmetry breaking, and a fractional time translation symmetry then appears. This can be seen in Supplementary Fig. \ref{phase for Fractional DTC}(a), in which the response for $\rho_{R_2R_2}$ goes through the phase of 2-DTC and its bifurcation into a series of stable fraction DTCs in the subharmonic regime. In this scenario, the discrete time translation symmetry of the phase of the 2-DTC is broken into numerous fractional $\mathbb{Z}_n$ symmetries [see the visible fractional $n$-DTCs with $n$ = 2.25, 2.36, and 2.42 in Supplementary Fig. \ref{phase for Fractional DTC}(a)]. These theoretical results are consistent with previously reported theories \cite{matus2019fractional,pizzi2021higher}.

\section*{SUPPLEMENTARY NOTE 3: DTCs versus RF frequency and intensity}

To validate the role of RF-field in our model, we compare the theoretical simulations and experimental observations. We calculate the phase map of the population $\rho_{R_2R_2}$ versus detuning $\delta$ (which corresponds to the frequency of RF field) and Rabi frequency $\Omega$ (which has relation to the intensity of RF-field), as given in Supplementary Fig. \ref{versus RF parameters}(a) and Supplementary Fig. \ref{versus RF parameters}(c). From Supplementary Fig. \ref{versus RF parameters}(a), we find that the order of DTCs changes against different detuning $\delta$ intervals, with similar behaviour to the experimental results given in Supplementary Fig. \ref{versus RF parameters}(b). By increasing Rabi frequency  $\Omega$, the system changes from 2-DTC to no-DTC, see the theoretical and experimental results in Supplementary Fig. \ref{versus RF parameters}(c) and Supplementary Fig. \ref{versus RF parameters}(d). The consistence between theoretical prediction and the experimental observations confirms effectiveness of our model.

\section*{SUPPLEMENTARY NOTE 4: RF-field induced shift}

When the frequency of the applied RF electric field is lower than $2\pi\times$10 MHz, the interaction between the RF electric field and the Rydberg atoms is nonresonant, and this condition results in an energy shift. For an RF field of $E(t) = E_{\rm{RF}} \rm{sin}(\omega t)$ with frequency $\omega$, the energy shift is $\Delta E =-\frac{1}{2}\alpha E^2(t)$, where $\alpha$ is the polarizability of the Rydberg states. We consider the time-averaged value of the energy shift and if we ignore the rapid oscillations of the RF electric field, we can then obtain:
\begin{equation}
\left \langle \Delta E \right \rangle = \left \langle -\frac{1}{2}\alpha E_{\rm{RF}}^2 \rm{sin}^2(\omega t) \right \rangle = -\frac{1}{4}\alpha E_{\rm{RF}}^2
\end{equation}
This energy shift causes an increase in the detuning between the Rydberg energy level and the ground state. Therefore, when an external RF electric field is applied, the detuning will turn into $\Delta_0 = -\frac{1}{4}\alpha E_{\rm{RF}}^2$. 

In the experiment, we scanned the electric field strength within the range from $0\sim1.75$ V/cm and obtained the three-photon electromagnetically-induced transparency (EIT) spectrum of the system, as shown in Supplementary Fig. \ref{shift}. The EIT spectral peak shifts gradually with increasing applied electric field strength. The dashed black line is the fitting function, which shows that the shift in the peak EIT satisfies a square relationship with the electric field strength, as calculated theoretically. In the experiment, the electric field just causes an ac Stark shift of energy level as the intensity of rf-field is small and the sideboards can be ignored, which leads to a shift of the EIT resonance peak.

\section*{SUPPLEMENTARY NOTE 5: Melting effect of 4-DTC}

In the experiment, we observed a gradual disappearance of the higher-order DTCs at a small driving frequency $f_0$, e.g., $f_0 = 2\pi\times$10 kHz. For the results shown in Supplementary Fig. 2(a) in the main text, when the detuning $\Delta > -2\pi\times 10.8$ MHz, the system breaks the short-range temporal order into a long-range temporal order as the frequency of the first peak decreases. In Supplementary Fig. \ref{melting}(a)-(d), we can see that the first peak frequency decreases from 0.25$f_0$ to 0.138$f_0$ because the range of the temporal order becomes longer. The states in these longer-range temporal orders are not immune to the increase in $\Delta$ [see also Supplementary Fig. 2(a) in the main text], and thus this change does not result in robust DTCs.

In addition, the shorter-range temporal order is lost, as shown by the reduced magnitude of the second peak [indicated by the red dashed circles in Supplementary Fig. \ref{melting}(a-d)]. In Supplementary Fig. \ref{melting}(c), there are no peaks within the frequency range from $2\pi\times$2.5 kHz $\sim$ $2\pi\times$7.5 kHz, and this corresponds to the destruction of the temporal orderliness. These gradually disappearing parts of the subharmonic response to periodic driving are regarded as the signature of the melting of the DTC order \cite{zaletel2023colloquium}. In addition, the temporal orderliness is retrieved when the detuning $\Delta$ is increased further [see Supplementary Fig. 2(b) in main text]; in this scenario, the orderliness of the 4-DTC is reproduced. This process somehow shows the process of melting and solidification of the 4-DTC.

\begin{figure*}
\centering
\includegraphics[width=2.0\columnwidth]{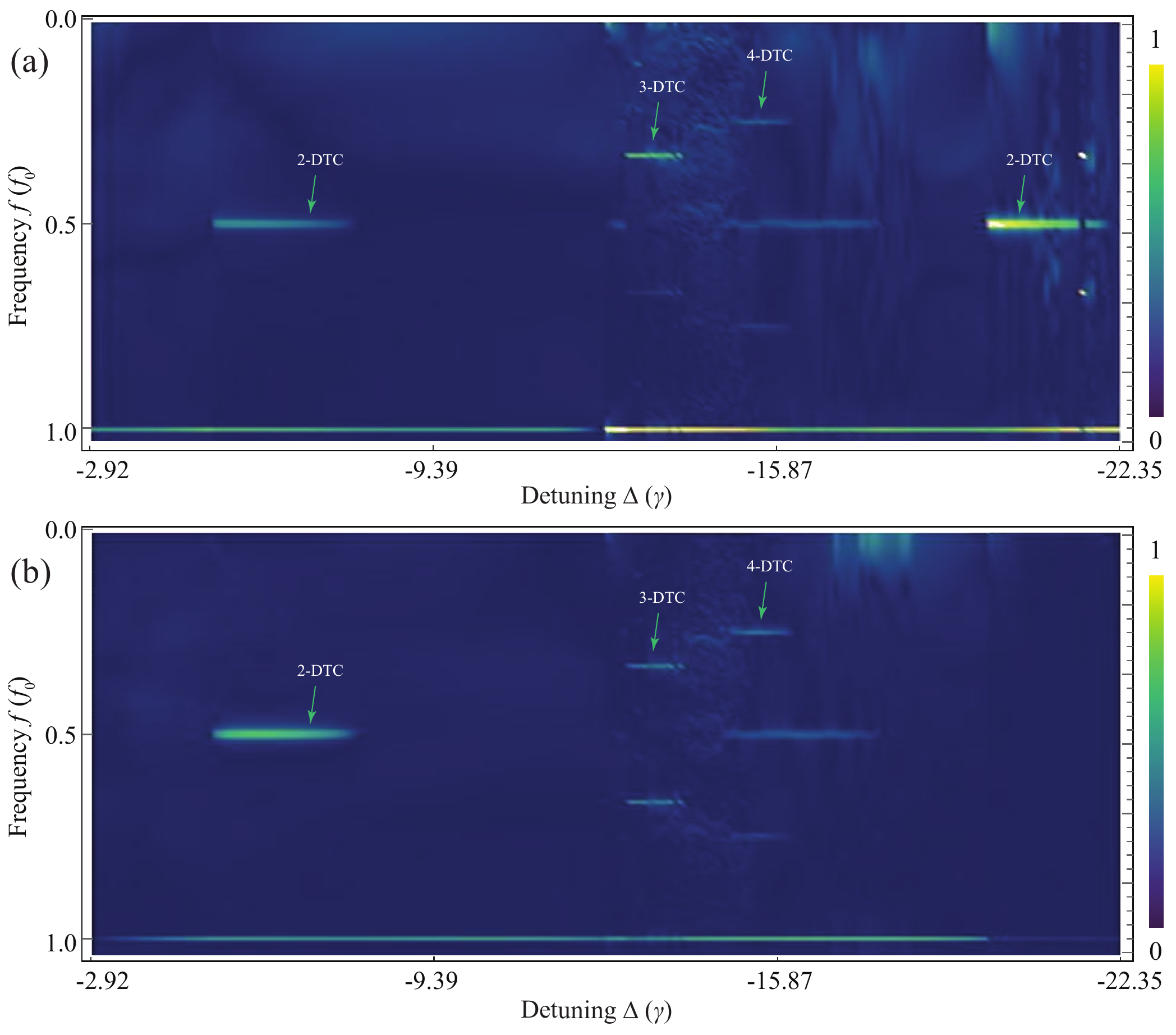}\\
\caption{\textbf{Theoretically simulated phase maps.} Calculated Fourier spectra of the Rydberg atom population $\rho_{R_1R_1}$ (a) and $\rho_{R_2R_2}$ (b) with parameters $V$ = -22.54$\gamma$, $\Omega$ = 5.62$\gamma$, $\delta$ = 11.69$\gamma$, and $\Delta_0$ = 9.72$\gamma$. In these results, the 2-DTC, 3-DTC, and 4-DTC phases are indicated using green arrows. The frequencies corresponding to these DTCs are $f$ = $f_0/2$, $f_0/3$, and $f_0/4$, where $f_0$ = 1.41$\gamma$ is the driving frequency. The color bar represents the Fourier transform intensity.}
\label{phase for 2-DTC}
\end{figure*}

\begin{figure*}
\centering
\includegraphics[width=2.0\columnwidth]{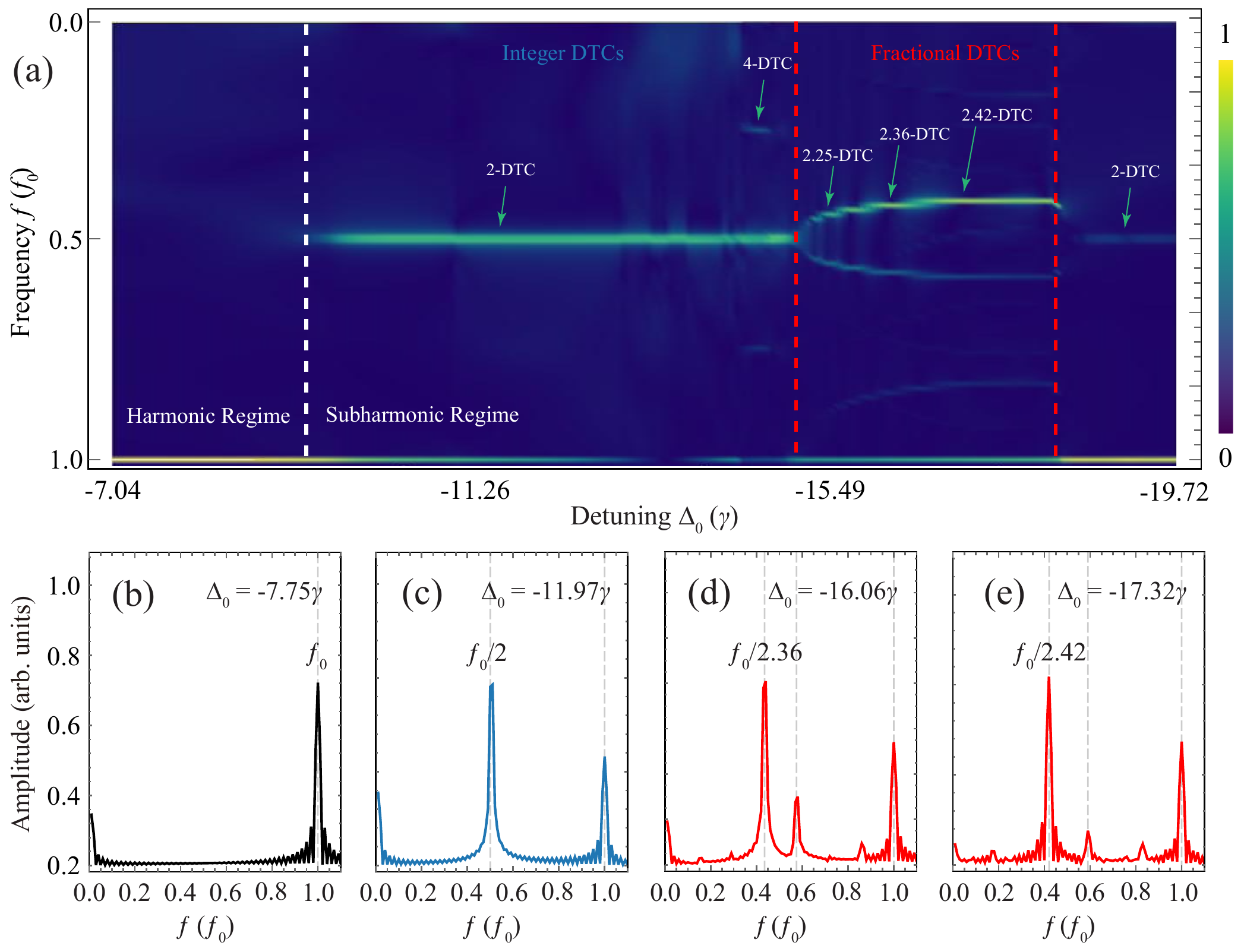}\\
\caption{\textbf{Bifurcation of 2-DTC and fractional DTCs.} (a) Calculated phase map of the population of the Rydberg atom $\rho_{R_2R_2}$ with parameters $V$ = -22.54$\gamma$, $\Omega_{1,2}$ = 5.63$\gamma$, $\delta$ = 10.99$\gamma$, and $\Delta$ = 0. The white dashed line separates the harmonic and subharmonic regimes. Here, $f_0$ = 1.41$\gamma$ is the driving frequency. The red dashed lines indicate the regime for appearance of the fractional DTCs. This theoretical phase diagram shows complex phases, including $n$-DTC phases with integers $n$ = 2 and 4, and several fractional DTCs. The color bar represents the Fourier transform intensity. (b) Fourier spectrum in the harmonic regime. (c)-(e) Fourier spectra, where the first peaks correspond to the phases of $n$-DTC with $n$ = 3, $n$ = 2.36, and $n$ = 2.42, respectively.}
\label{phase for Fractional DTC}
\end{figure*}

\begin{figure*}
\centering
\includegraphics[width=2.0\columnwidth]{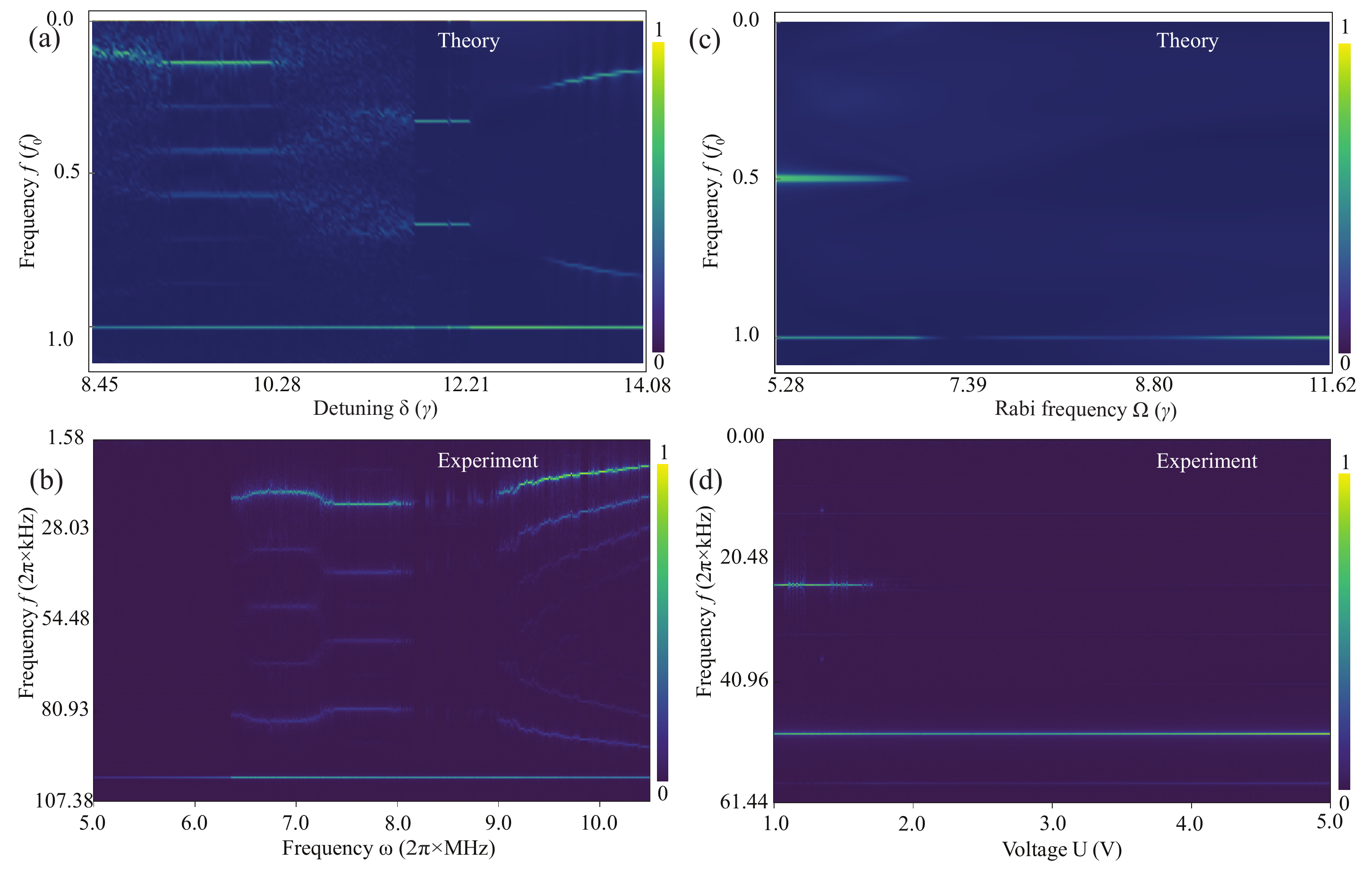}\\
\caption{\textbf{DTCs versus RF frequency and intensity.} Calculated phase map of the population of the Rydberg atom $\rho_{R_2R_2}$ versus detuning $\delta$ (a) and Rabi frequency $\Omega$ (c). For simulating results in (a), we set the parameters $V$ = -22.54$\gamma$, $\Omega$ = 5.28$\gamma$; For simulating results in (b), we set the parameters $V$ = -22.54$\gamma$, $\delta$ = 10.42$\gamma$. Measured phase map versus the frequency of RF-field $\omega$ (b) [with parameters of $U=1.2$ V and $f_0=100$ kHz] and the voltage of RF-field $U$ (d) [with $\omega=2\pi\times7.1$ MHz and $f_0=50$ kHz].}
\label{versus RF parameters}
\end{figure*}

\begin{figure*}
\includegraphics[width=2\columnwidth]{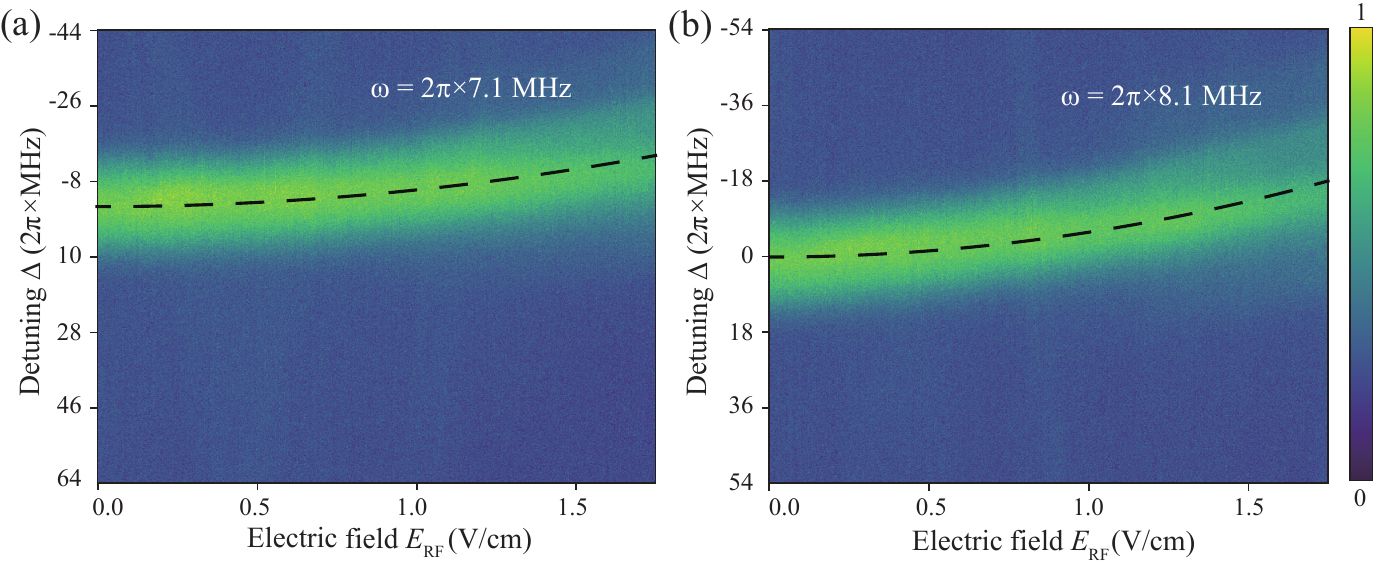}\\
\caption{\textbf{RF field-induced shift.} EIT spectra of the Rydberg atoms in the $0\sim1.75$ V/cm electric field strength range, at the frequencies of $\omega = 2\pi\times7.1$ MHz in (a) and $\omega = 2\pi\times8.1$ MHz in (b).}
\label{shift}
\end{figure*}

\begin{figure*}
\centering
\includegraphics[width=2.0\columnwidth]{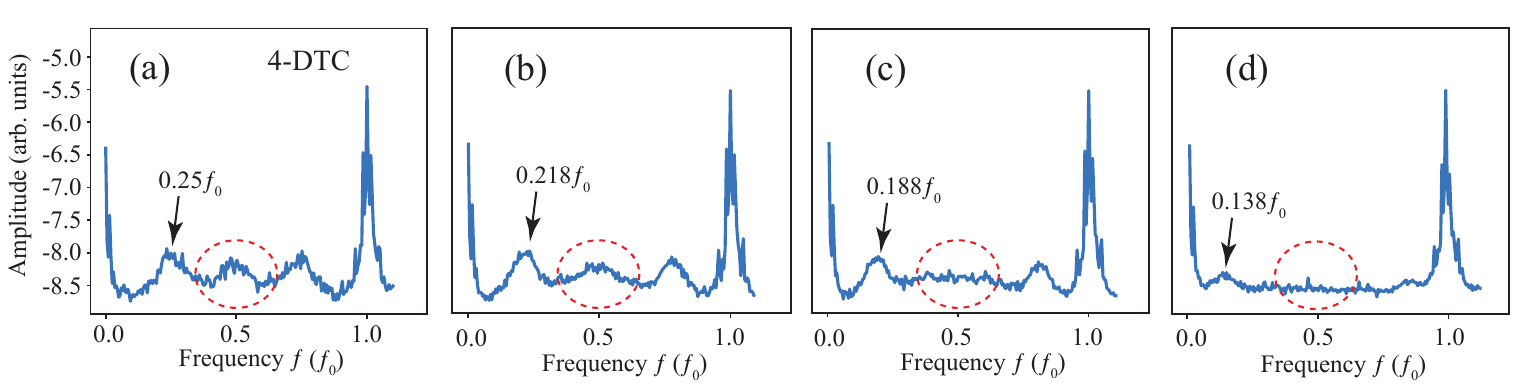}\\
\caption{\textbf{Melting effect.} (a)-(d) Measured Fourier spectra at $\Delta = -2\pi\times10.8$ MHz $\sim$ $\Delta = -2\pi\times 8.65$ MHz. The frequency of the RF field is $f_0 = 2\pi\times$10 kHz. (a) represents the higher-order $n$-DTC with $n$ = 4. (b)-(d) are the measured Fourier spectra, where the first peaks have frequencies of 0.218$f_0$, 0.188$f_0$, and 0.138$f_0$, respectively. The red dashed circles illustrate the missing frequency peaks. The data presented here were measured from 200 multiple experimental trials for high visibility.}
\label{melting}
\end{figure*}

\end{document}